\documentclass[12pt]{revtex4}

\usepackage{epsf}
\usepackage{graphics}
\usepackage{epsfig}
\usepackage{amsmath}  
\usepackage{amsfonts} 
\usepackage{amssymb}  

\begin{document}

\title{Spin-glass behaviour on random lattices}

\author{M O Hase$^{1, 2, a}$, J R L de Almeida$^{3, b}$ and S R Salinas$^{2, c}$}

\affiliation{
$^{1}$Escola de Artes, Ci\^encias e Humanidades, Universidade de S\~ao Paulo, Avenida Arlindo B\'ettio 1000, 03828-000, S\~{a}o Paulo, SP, Brazil\\
$^{2}$Instituto de F\'isica, Universidade de S\~ao Paulo, Caixa Postal 66318 05315-970, S\~ao Paulo, SP, Brazil\\
$^{3}$Departamento de F\'isica, Universidade Federal de Pernambuco, 50670-901, Recife, PE, Brazil\\
$^{a}$mhase@usp.br, $^{b}$almeida@df.ufpe.br and $^{c}$ssalinas@if.usp.br}

\begin{abstract}
The ground-state phase diagram of an Ising spin-glass model on a random graph with an arbitrary fraction $w$ of ferromagnetic interactions is analysed in the presence of an external field. Using the replica method, and performing an analysis of stability of the replica-symmetric solution, it is shown that $w=1/2$, correponding to an unbiased spin glass, is a singular point in the phase diagram, separating a region with a spin-glass phase ($w<1/2$) from a region with spin-glass, ferromagnetic, mixed, and paramagnetic phases ($w>1/2$).
\end{abstract}
\pacs{75.10.Nr, 75.50.Lk, 89.75.-k}

\maketitle



\section{Introduction}

The mean-field formulations of the Ising spin glass\cite{SK75} (an interesting introduction to spin-glass theory based on $p$-spin spherical model is found in \cite{CC05}) and related disordered models on a lattice are known to lead to replica - symmetry breaking, which is in turn associated with the existence of a large number of metastable states. The number of local minima in the free energy landscape of these solutions grows exponentially with the size of the lattice\cite{BY86, MPV87}. The impressive richness of these mean-field results has been a motivation to look at more realistic spin-glass models, with the inclusion of the effects of either short-range interactions or the finite coordination of the crystal lattices. Besides the implementation of the Ising spin glass (SG) on a Bethe lattice\cite{MP01, PDH02, CKRT05, K05}, which is one of the natural ways to deal with a finite connectivity, there is a work by Viana and Bray\cite{VB85} for the Ising SG on a random graph with a mean connectivity. According to a number of investigations\cite{KS87, MP87, MdD87, dAdDM88}, the model of Viana and Bray (VB), which is tractable by the replica method, leads to a glassy phase in the ground state. Finitely connected disordered models\cite{dAB87, PPR03, JKK08} usually lead to a set of self-consistent equations that can be examined by exact numerical methods\cite{MP01}, while some analytical results can be achieved for special situations\cite{HdAS05}.

The problem of a statistical model in a finitely connected graph has been mostly investigated by numerical methods\cite{WNH05, BSLW06, MS06, HPV08, H08, H09, ABS10}. Critical phenomena associated with these systems are known to display a much richer behaviour than that predicted by mean-field calculations\cite{DGM08}. It may be remarked that models with finite connectivity are closer to ``realistic'' Bravais lattices than the analogous mean-field versions. In this article, some results for the ground state of a frustrated $\pm J$ Ising spin glass on a random graph, in the presence of an external magnetic field, and with an arbitrary fraction of ferromagnetic interactions are reported. A parameter $w$ that gauges the concentration of these ferromagnetic ($J_{0}>0$) interactions is introduced. For $w=1/2$, this model is a pure, unbiased, spin glass; ferromagnetic (antiferromagnetic) interactions are favoured for $w>1/2$ ($w<1/2)$. It is known that the presence of an external field always increases the technical complexity of the solutions of these problems. In this analysis, the phase diagram in the ground state for external fields that are multiples of $J_{0}$ is obtained; this resembles a choice of random fields that has already been adopted in a previous work \cite{HdAS05}. The absence of external fields leads to the problem of colouring graphs, which can be treated, from the statistical physics standpoint, as the problem of investigating the ground state of a diluted Potts antiferromagnet\cite{MPWZ02, ZK07}. The topology of the random network favours the appearance of a spin-glass phase for $w<1/2$. For $w\geq 1/2$, the usual spin-glass, ferromagnetic, mixed and paramagnetic phases are obtained. It should be emphasized that even in the presence of overall antiferromagnetic interactions ($w<1/2$) the inherent disorder of the lattice still yields a spin-glass phase. This behavior is related to the type of graph examined in this work, which belongs to the Erd\"os-R\'enyi type without sublattices. Therefore, despite the fact that the model examined in this paper is a diluted antiferromagnet in a uniform magnetic field, there is no (known) connection with the random-field Ising model\cite{FA79, C84}, which has been shown in a recent work\cite{KRTZ10} to display no spin-glass phase for Ising spins (the continuous spin version of this result is found in \cite{KRTSZ11}).


\section{Definition of the model}

Consider the Hamiltonian

\begin{eqnarray}
\mathcal{H}(\{\sigma_{x}\}) = -\sum_{(x, x^{\prime})\in\Lambda_{N}} J_{x, x^{\prime}} \sigma_{x} \sigma_{x^{\prime}} - H \sum_{x\in\Lambda_{N}} \sigma_{x}\,,
\label{Ham}
\end{eqnarray}

\noindent
where $\sigma_{x}\in\{-1, 1\}$ is an Ising spin on site $x$, belonging to a finite lattice $\Lambda_{N}$ of $N$ sites and $H$ is the external magnetic field. The first sum is over all distinct pairs of lattice sites. The exchange integrals $\{J_{x,x^{\prime }}\}$ are independent and identically distributed random variables, associated with the distribution

\begin{eqnarray}
p_{J}(J_{x, x^{\prime}}) = \left(1 - \frac{c}{N}\right) \delta(J_{x, x^{\prime}}) + \frac{c}{N} \rho(J_{x, x^{\prime}})\,,
\label{pJ}
\end{eqnarray}

\noindent
where 

\begin{eqnarray}
\rho(J_{x, x^{\prime}}) = w \delta(J_{x, x^{\prime}} - J_{0}) + (1 - w) \delta(J_{x, x^{\prime}} + J_{0})\,,
\label{rho}
\end{eqnarray}

\noindent
and $J_{0}$ is a positive parameter. According to the distribution (\ref{pJ}), two sites are connected (disconnected) with a probability $c/N$ ($1-c/N$), where $c$ is the mean connectivity of the lattice. In this diluted $\pm J$ model, parameter $w$ weighs the fraction of ferromagnetic bonds.

\section{Replica-symmetric solution}

Using the replica method, the variational free energy is written as

\begin{eqnarray}
\nonumber f[G] & = & \frac{1}{\beta} \lim_{n\rightarrow 0} \left\{ \frac{c}{2} + \frac{c}{2} \sum_{r=0}^{n} \sum_{(\alpha_{1}, \cdots, \alpha_{r})} b_{r} q_{\alpha_{1}, \cdots, \alpha_{r}}^{2} - \right. \\
 & & \left. - \ln \mbox{Tr}_{\{\sigma\}} \exp \left[ G(\{\sigma_{\alpha}\}) + \beta H\sum_{\alpha = 1}^{n} \sigma_{\alpha} \right] \right\}\,,
\label{fvar}
\end{eqnarray}

\noindent
where

\begin{eqnarray}
b_{r} = \int\limits_{\mathbb{R}} dJ \rho(J) \cosh^{n}(\beta J)\tanh^{r}(\beta J)\,,
\label{br}
\end{eqnarray}

\noindent
and

\begin{eqnarray}
G(\{\sigma_{\alpha}\}) = c \sum_{r=0}^{n} \sum_{(\alpha_{1}, \cdots, \alpha_{r})} b_{r} q_{\alpha_{1}, \cdots, \alpha_{r}} \sigma_{\alpha_{1}} \cdots \sigma_{\alpha_{r}}
\label{G}
\end{eqnarray}

\noindent
is the global order parameter\cite{MdD87, dDM87}, which takes into account the $2^{n}$ order parameters of this problem (in contrast to the Sherrington - Kirkpatrick (SK) model\cite{SK75}, which has just two order parameters).

The stationary free energy comes from the minimization of the variational free energy (\ref{fvar}) with respect to the set $\{q_{\alpha _{1},\cdots,\alpha _{r}}\}$, which leads to the stationary conditions

\begin{eqnarray}
q_{\alpha_{1}, \cdots, \alpha_{r}} = \frac{ \mbox{Tr}_{\{\sigma\}} \sigma_{\alpha_{1}} \cdots \sigma_{\alpha_{r}} \exp \left[ G(\{\sigma_{\alpha}\}) + \beta H \sum_{\alpha = 1}^{n} \sigma_{\alpha} \right] }{ \mbox{Tr}_{\sigma} \exp \left[ G(\{\sigma_{\alpha}\}) + \beta H \sum_{\alpha = 1}^{n} \sigma_{\alpha}\right] }\,.
\label{qst}
\end{eqnarray}

\indent

The replica-symmetric (RS) solution of this problem may be obtained from the introduction of a local field $h$, with an effective probability distribution $P(h)$, which is equally applied to all of the replica spin variables. Thus, one has

\begin{eqnarray}
q_{\alpha_{1}, \cdots, \alpha_{r}} = \int\limits_{\mathbb{R}} dh P(h) \tanh^{r}(\beta h)\,.
\label{qth}
\end{eqnarray}

\noindent
Moreover, in the replica-symmetric context, one can see that $G(\{\sigma\}) = G(\hat \sigma)$, where $\hat\sigma = \sum_{\alpha=1}^{n} \sigma_{\alpha}$.

Equation (\ref{qth}) leads to a relation between the distribution of local fields, $P$, and the global order parameter, $G$, which is

\begin{eqnarray}
G(iy/\beta) = c\int\limits_{\mathbb{R}}dJ\rho(J)\int\limits_{\mathbb{R}}dhP(h)e^{\frac{iy}{\beta}\tanh^{-1}\left[\tanh(\beta J)\tanh(\beta h)\right]}\,.
\label{GP}
\end{eqnarray}

The analysis will be carried out at the ground state ($\beta\rightarrow\infty$). Also, in analogy to a previous work\cite{HdAS05}, the RS solutions of this problem are restricted to the condition $r=H/J_{0}$, where $r$ is an integer. In other words, the solutions in this work are written in the form

\begin{eqnarray}
P(h) = \sum_{k\in\mathbb{Z}} a_{k} \delta(h - kJ_{0})\,,
\label{pdisc}
\end{eqnarray}

\noindent
which takes into account the discrete part of the distribution function only; the coefficients $\{a_{k}\}$ are given below. Under these conditions, by replacing (\ref{pdisc}) in (\ref{GP}), one can show that

\begin{eqnarray}
\tilde G(iy) = \lim_{\beta\rightarrow\infty} G(iy/\beta) = A + B^{\,\prime} e^{iyJ_{0}} + C^{\,\prime} e^{-iyJ_{0}}\,,
\label{tildeG}
\end{eqnarray}

\noindent
where

\begin{eqnarray}
\nonumber A = ca_{0} \quad , \quad B^{\,\prime} = wB + (1 - w)C \quad , \quad C^{\,\prime} = (1 - w)B + wC\,, \\
\end{eqnarray}

\noindent
with

\begin{eqnarray}
\nonumber B = c\sum_{k=1}^{\infty} a_{k}\,, \;\; C = c\sum_{k=-\infty}^{-1} a_{k} \;\; \mbox{and} \;\; a_{k} = e^{A - c} \left(\frac{B^{\,\prime}}{C^{\,\prime}}\right)^{\frac{k - r}{2}} I_{k-r}(2\sqrt{B^{\,\prime}C^{\,\prime}})\,, \\
\label{ABCBCak}
\end{eqnarray}

\noindent
where $I_{\nu}(x)$ is the modified Bessel function of order $\nu$.

The solution of this problem depends on three equations, from which one can obtain $A$, $B$ and $C$. Using the previous equations, one finds

\begin{eqnarray}
A = c e^{A-c}\left(\frac{C^{\,\prime}}{B^{\,\prime}}\right)^{\frac{r}{2}} I_{r}(2\sqrt{B^{\,\prime}C^{\,\prime}})\,, \quad c = A + B^{\,\prime} + C^{\,\prime} \; \left(\,= A + B + C \right)
\label{AC}
\end{eqnarray}

\noindent
and

\begin{eqnarray}
\nonumber B^{\,\prime} & = & ce^{A-c} \bigg[ we^{c-A} - we^{C^{\,\prime}}\delta_{r, 0} - \\
\nonumber & & - wC^{\,\prime}\left(\frac{C^{\,\prime}}{B^{\,\prime}}\right)^{\frac{r-1}{2}} \int\limits_{0}^{1}du e^{C^{\,\prime}u}\left(1-u\right)^{\frac{r-1}{2}}I_{r-1}\left(2\sqrt{B^{\,\prime}C^{\,\prime}(1-u)}\right) + \\
\nonumber & & + \left(1-w\right)C^{\,\prime}\left(\frac{C^{\,\prime}}{B^{\,\prime}}\right)^{\frac{r}{2}} \int\limits_{0}^{1}du e^{C^{\,\prime}u}\left(1-u\right)^{\frac{r}{2}}I_{r}\left(2\sqrt{B^{\,\prime}C^{\,\prime}(1-u)}\right) \bigg]\,. \\
\label{B}
\end{eqnarray}


\section{Analysis of stability}

The stability analysis can be performed using the same steps as previous works\cite{MdD87, HdAS05}. It is based on the search for the eigenvalues of the Hessian matrix associated with the variational free energy. By some technical manipulations of the eigenvalue equations, which are similar to previous calculations\cite{dDM87, HdAS05}, a separate analysis for the cases $w\neq 1/2$ and $w=1/2$ is carried out.

For $w\neq 1/2$, one finds the following transverse eigenvalues:



\begin{eqnarray}
\nonumber \lambda & = & \frac{1}{c}\,,\;\; - \frac{A}{c} + \frac{1}{c}\,,\;\; - \frac{A}{c} - \frac{1}{c(1 - 2w)} \;\; \mbox{and} \;\; - \frac{A}{c} - \frac{1}{1 - 2w}\left(\frac{w}{c}+\Delta^{\pm}\right)\,, \\
\label{lambda}
\end{eqnarray}

\noindent
where $\Delta^{\pm} = \pm \sqrt{ \left[ \frac{1 - w}{c} \pm a_{1}\left(1 - 2w\right) \right] \left[ \frac{1 - w}{c} \pm a_{-1}\left(1 - 2w\right) \right]}$. The stability of the RS solution depends on the smallest eigenvalue,

\begin{eqnarray}
\lambda_{0} = - \frac{A}{c} - \frac{w}{c\left(1 - 2w\right)} - \frac{1}{1 - 2w} \times \left\{
\begin{array}{lcl}
\Delta^{-} & , & w>\frac{1}{2} \\
\Delta^{+} & , & w<\frac{1}{2}
\end{array}
\right.\,.
\label{lambda0}
\end{eqnarray}

\begin{figure}[htp]
{\includegraphics[scale=0.45]{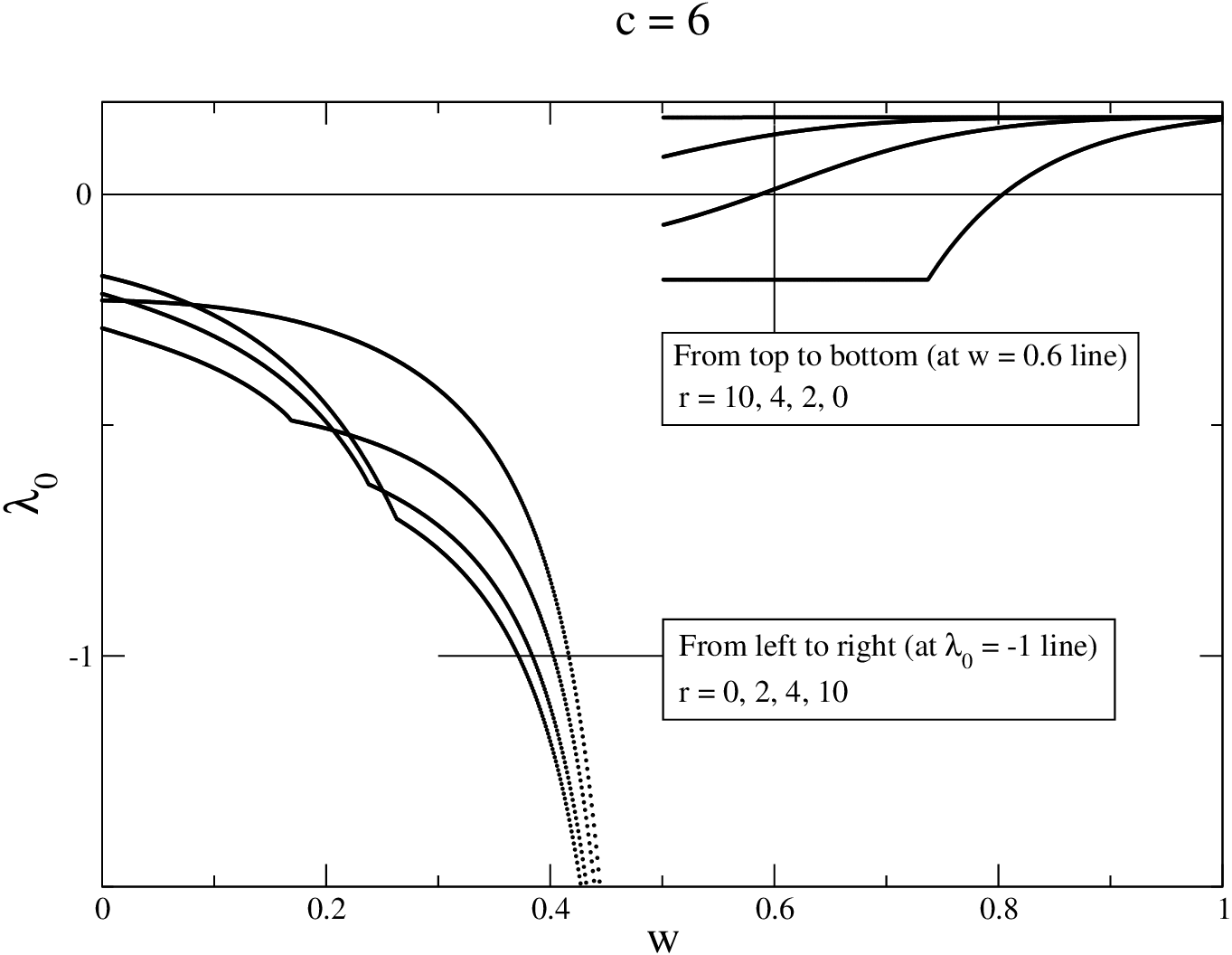}}
\caption{Graph of $\lambda_{0}$ \textit{versus} $w$, obtained by numerical calculation of the eigenvalue.}
\end{figure}

The graph of $\lambda_{0}$ \textit{versus} $w$ is shown in figure 1 for $c=6$. Some limiting cases are analytically accessible. In the pure ferromagnetic case ($w=1$), one has $\lambda _{0}(w=1)=-\frac{A}{c}+\frac{1}{c}-\sqrt{a_{1}a_{-1}}$, which confirms the stability of the RS solution in large external fields, in which case $\lambda _{0}(w=1,H\sim \infty )\sim 1/c$ (since, for a \textit{fixed} value of $k$, $a_{k}$ goes to zero for sufficiently large $H$). Moreover, by the same argument, one can make $\lambda _{0}$ positive for any $w>1/2$ if the field is sufficiently large. On the other hand, the $w=0$ (pure antiferromagnetic interactions) case implies $\lambda _{0}(w=0)=-\frac{A}{c}-\sqrt{\left[ \frac{1}{c}+a_{1}\right]\left[ \frac{1}{c}+a_{-1}\right] }$, which is clearly negative. This means that the RS solution for this finitely connected spin-glass model is always unstable in this regime, even in the limit $H\rightarrow \infty $, which yields $\lambda _{0}(w=0,H\rightarrow \infty )=-1/c$. Note that in the limit of infinite connectivity, $c\rightarrow \infty $, this eigenvalue goes to zero, becoming marginally stable as in the SK model. The instability of the RS solution even for large external fields makes the Viana-Bray model distinct from the Bethe lattice. While the latter has \textit{fixed} connectivity, the former has fixed \textit{mean} connectivity, and the degree distribution follows a Poisson distribution. This means that there are vertices that have infinite degree in the thermodynamic limit, and the signs of those vertices are determined not by the external field, but the interaction with their neighbours only. Like the SK model, the external field is not able to stabilize the RS solution\cite{dAT78} in the ground state in the Viana-Bray lattice. On the other hand, the whole region where the external field is large is dominated by the RS solution if the system is analysed on a Bethe lattice\cite{JKK08}. The entropy os the model, also based on the context of RS solution and discrete field (\ref{pdisc}), is evaluated in the appendix of this work, although no precise statements can be made in the $w<1/2$ region.

\begin{figure}[htp]
{\includegraphics[scale=0.45]{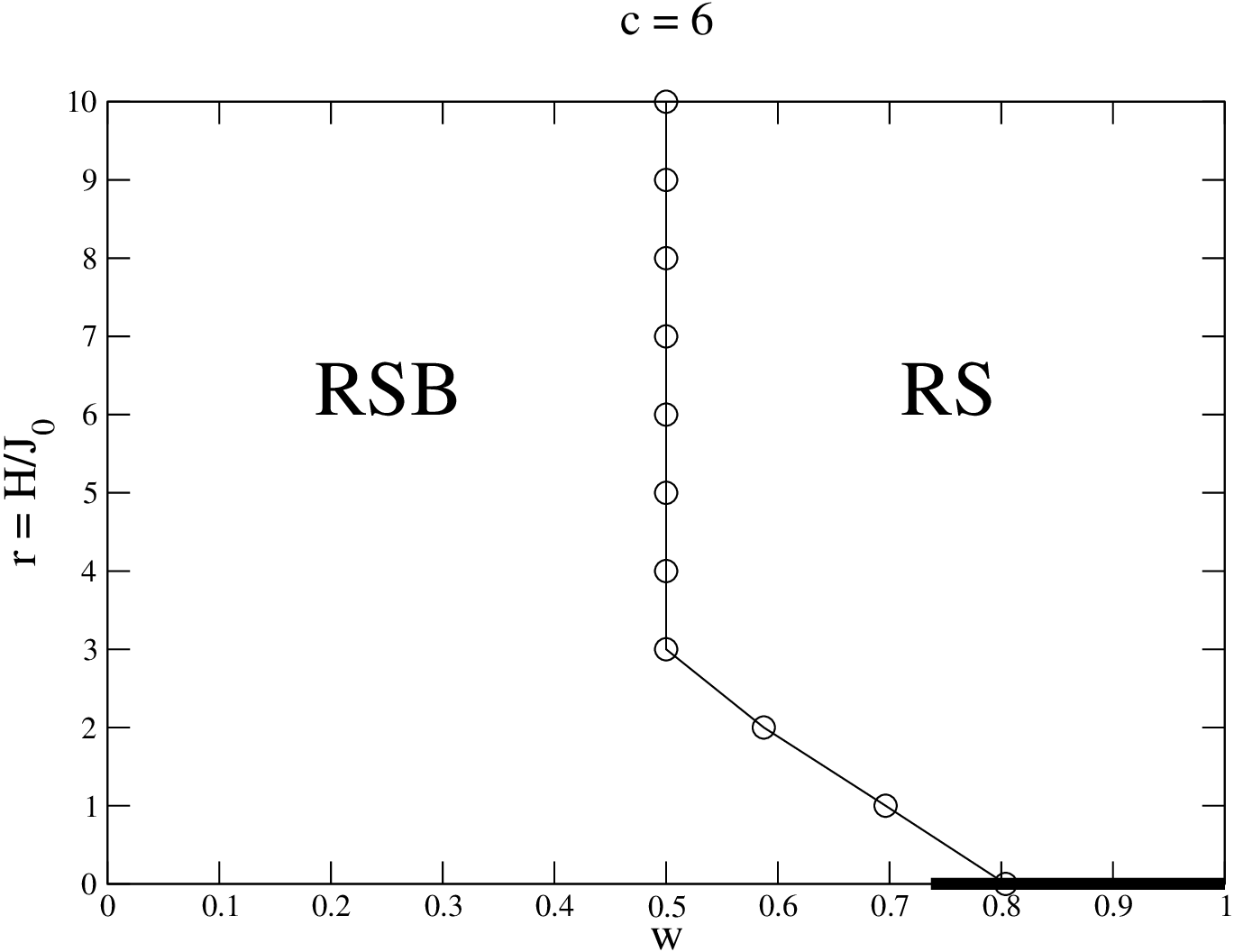}}
\caption{The replica-symmetry-breaking (RSB) and RS phases in the $r-w$ phase diagram. The thick line at $r=0$ indicates spontaneous magnetization.}
\end{figure}

The phase diagrams (figure 2), which are qualitatively similar for other choices of the mean connectivity $c$, show that the RS solution is always unstable for $w<1/2$, which can also be seen from equation (\ref{lambda0}). For $w>1/2$, the RS solution is stable above a critical value $w^{\ast }$ of the concentration, which depends on the applied field. This behaviour can be illustrated in the $r-w$ phase diagram, where the SG phase is separated from the paramagnetic phase by a de Almeida-Thouless line\cite{dAT78}.

In zero field one can show the presence of ferromagnetic and mixed phases. This mixed phase is characterized by the asymmetric distribution of the local field, which leads to $a_{k}\neq a_{-k}$ for some values of $k\in \mathbb{Z}$.

For $w=1/2$, the eigenvalues of the stability matrix are

\begin{eqnarray}
\lambda & = & \frac{1}{c}\,, \quad - \frac{A}{c} + \frac{1}{c} \quad\mbox{and} \quad - \frac{A}{c} + \frac{1}{c} \pm \frac{1}{2}\left( a_{1} + a_{-1} \right)\,.
\label{lambda1/2}
\end{eqnarray}

\noindent
The smallest of these eigenvalues is given by

\begin{eqnarray}
\lambda_{0} = - \frac{A}{c} + \frac{1}{c} - \frac{1}{2} \left(a_{1} + a_{-1}\right)\,.
\label{lambda01/2}
\end{eqnarray}

\noindent
This eigenvalue does not have a definite sign, but it is positive (tending to $1/c$) for sufficiently large field, and agrees with equation (\ref{lambda0}) when $w\downarrow 1/2$ ($\lambda_{0}(w)$ is continuous from the right). On the other hand, it is easy to see that $\lim_{w\uparrow 1/2}\lambda_{0}(w)=-\infty$.


\section{Conclusions}

In summary, this work reports a calculation of the phase diagram in the ground state of a diluted $\pm J$ Ising spin-glass model, in an external field, with a fraction $w$ of ferromagnetic bonds. For $w<1/2$, which corresponds to a larger concentration of antiferromagnetic bonds, the topology of the lattice leads to the existence of a spin-glass phase only. It should be remarked that this behaviour has also been found in numerical calculations for antiferromagnetic Ising models on free-scale graphs\cite{H09, BSLW06} and on small-world networks\cite{H08}. For $w\geq 1/2$, with a larger concentration of ferromagnetic bonds, the usual phases are found.


\bigskip

The authors thank the Brazilian agencies CNPq and CAPES for financial support. MOH would like to thank the Departamento de F\'{\i}sica da Universidade Federal de Pernambuco (DF-UFPE), where this work was completed, for their kind hospitality.


\section{Appendix}

In this appendix, the explicit form of the entropy in the replica-symmetric context is presented. For this purpose, it is convenient to express the variational free energy in terms of the local field distribution\cite{MP87} $P$ as
\begin{eqnarray}
\nonumber f[P](\beta) & = & -\frac{c}{2\beta}\int\limits_{\mathbb{R}}dJ\rho(J)\ln\big[\cosh(\beta J)\big] - \frac{1}{\beta}\int\limits_{\mathbb{R}}dhP(h)\ln\big[2\cosh(\beta J)\big]+ \\
\nonumber & & +\frac{c}{2\beta}\int\limits_{\mathbb{R}}dJ\rho(J)\int\limits_{\mathbb{R}}dhP(h)\int\limits_{\mathbb{R}}dh^{\prime}P(h^{\prime})\ln\big[1+\tanh(\beta J)\tanh(\beta h)\tanh(\beta h^{\prime})\big]-\\
 & & -\frac{c}{2\beta}\int\limits_{\mathbb{R}}dJ\rho(J)\int\limits_{\mathbb{R}}dhP(h)\ln\big[1-\tanh^{2}(\beta J)\tanh^{2}(\beta h)\big]\,.
\label{varF}
\end{eqnarray}
The entropy in the ground state ($\beta\rightarrow\infty$), $s_{0}$, is then
\begin{eqnarray}
\nonumber s_{0}/k_{B} & = & a_{0}\ln 2-\frac{c}{2}\left(1-a_{0}\right)^{2}\ln 2 - \frac{c}{2}\Bigg\{ \ln 2 - \left(1-a_{0}\right)\ln 4 - \left(a_{1}+a_{-1}\right)\ln 2 + \\
\nonumber & & +w\left[ \left(\frac{a_{1}C+a_{-1}B}{c}\right)\ln 4 + 2a_{1}a_{-1}\ln\left(\frac{3}{4}\right)\right] + \\
 & & +\left(1-w\right)\left[ \left(\frac{a_{1}B+a_{-1}C}{c}\right)\ln 4 + \left(a_{1}^{2}+a_{-1}^{2}\right)\ln\left(\frac{3}{4}\right)\right] \Bigg\}\,,
\label{s}
\end{eqnarray}
which recovers the entropy obtained in \cite{dA03} for $w=1/2$. A graph $s/k_{B}$ \textit{versus} $w$ is provided in figure 3 for $c=6$.

\begin{figure}[htp]
{\includegraphics[scale=0.45]{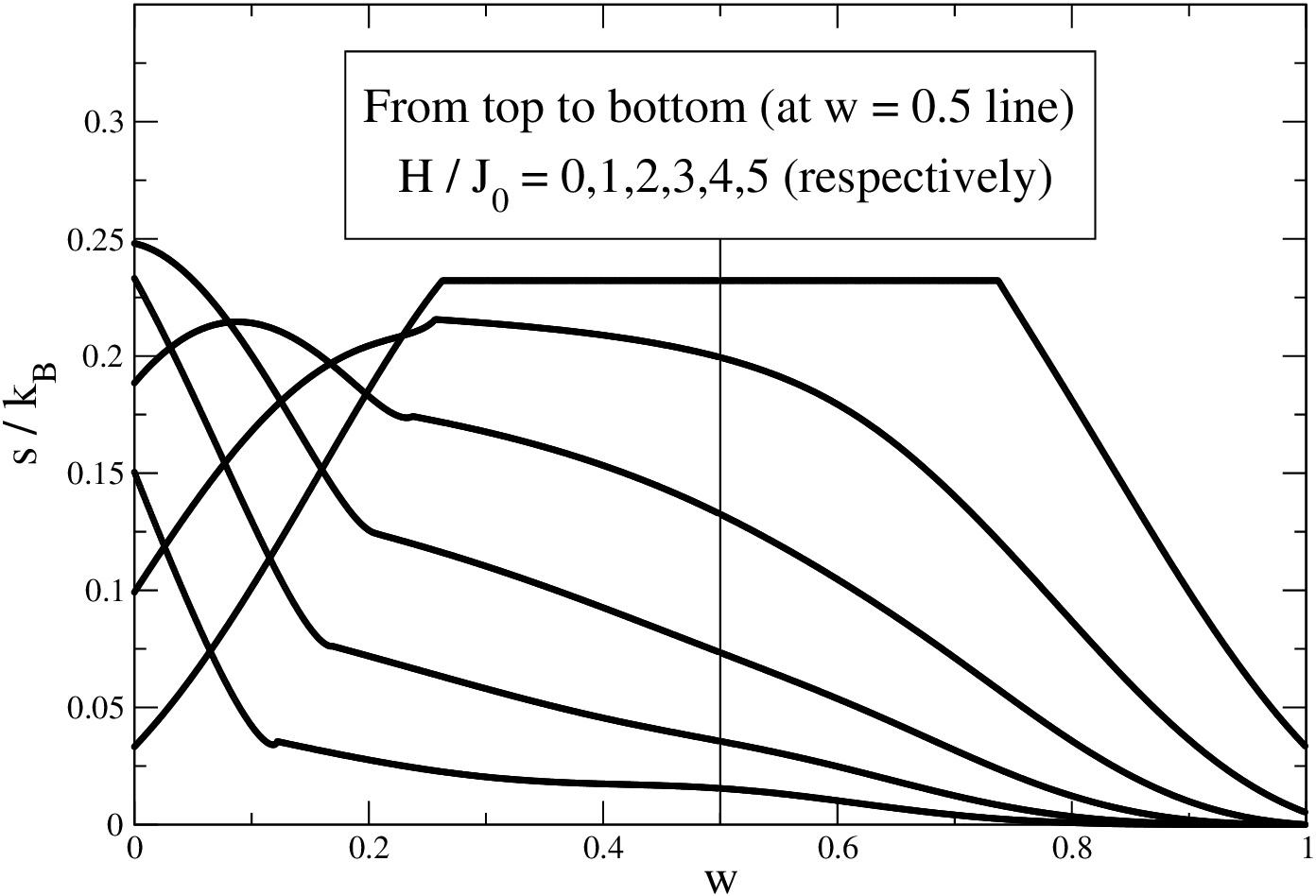}}
\caption{Graph $s/k_{B}$ \textit{versus} $w$.}
\end{figure}

Note that for large external field, the entropy above vanishes for any choice of $w$, which does not mean that a RS solution is expected for $w<1/2$. In the presence of an external field, the RS ferromagnetic limit ($w=1$) leads the entropy to zero, which shows that the discrete solution (\ref{pdisc}) has succeeded in capturing this expected feature even when not considering the continuous part of the field distribution $P$. Nevertheless, no precise statement can be made in the region $w<1/2$, where the RS solution fails. In this case, both the RSB analysis and a satisfactory form of the solution for the field distribution (that involves also the continuous part, which was ignored in this work in order to achieve analytical results) is required.



\begin{thebibliography}{999}

\bibitem{SK75} Sherrington D, Kirkpatrick S, \textit{Phys. Rev. Lett.} \textbf{35}, 1792 (1975)

\bibitem{CC05} Castellani T, Cavagna A, \textit{J. Stat. Mech.} P05012 (2005)

\bibitem{BY86} Binder K, Young A P, \textit{Rev. Mod. Phys.} \textbf{58}, 801 (1986)

\bibitem{MPV87} M\'ezard M, Parisi G, Virasoro M, \textit{Spin Glass Theory and Beyond} (World Sci., 1987)

\bibitem{MP01} M\'ezard M, Parisi G, \textit{Eur. Phys. J. B} \textbf{20}, 217 (2001)

\bibitem{PDH02} Pastor A A, Dobrosavljevi\'{c} V, Horbach M L, \textit{Phys. Rev. B} \textbf{66}, 014413 (2002)

\bibitem{CKRT05} Castellani T, Krzakala F, Ricci-Tersenghi F, \textit{Eur. Phys. J. B} \textbf{47}, 99 (2005)

\bibitem{K05} Krzakala F, \textit{Prog. Theor. Phys. Suppl.} \textbf{157}, 77 (2005)

\bibitem{VB85} Viana L, Bray A J, \textit{J. Phys. C} \textbf{18}, 3037 (1985)

\bibitem{KS87} Kanter I, Sompolinsky H, \textit{Phys. Rev. Lett.} \textbf{58}, 164 (1987)

\bibitem{MP87} M\'{e}zard M, Parisi G, \textit{Europhys. Lett.} \textbf{3}, 1067 (1987)

\bibitem{MdD87} Mottishaw P, DeDominicis C, \textit{J. Phys. A} \textbf{20}, L375 (1987)

\bibitem{dAdDM88} de Almeida J R L, DeDominicis C, Mottishaw P, \textit{J. Phys. A} \textbf{21}, L693 (1988)

\bibitem{dAB87} de Almeida J R L, Bruinsma R, \textit{Phys. Rev. B} \textbf{35}, 7267 (1987)

\bibitem{PPR03} Pagnani A, Parisi G, Rati\'eville, \textit{Phys. Rev. E} \textbf{68}, 046706 (2003)

\bibitem{JKK08} J\"org T, Katzgraber H G, Krzakala F, \textit{Phys. Rev. Lett.} \textbf{100}, 197202 (2008)

\bibitem{HdAS05} Hase M O, de Almeida J R L, Salinas S R, \textit{Eur. Phys. J. B} \textbf{47}, 245 (2005)

\bibitem{WNH05} Wemmenhove B, Nikoletopoulos T, Hatchett J P L, \textit{J. Stat. Mech.} P11007 (2005)

\bibitem{BSLW06} Bartolozzi M, Surungan T, Leinweber D B, Williams A G, \textit{Phys. Rev. B} \textbf{73}, 224419 (2006)

\bibitem{MS06} Migliorini G, Saad D, \textit{Phys. Rev. E} \textbf{73}, 026122 (2006)

\bibitem{HPV08} Hasenbusch M, Pelissetto A, Vicari E, \textit{Phys. Rev. B} \textbf{78}, 214205 (2008)

\bibitem{H08} Herrero C P, \textit{Phys. Rev. E} \textbf{77}, 041102 (2008)

\bibitem{H09} Herrero C P, \textit{Eur. Phys. J. B} \textbf{70}, 435 (2009)

\bibitem{ABS10} Agliari E, Burioni R, Sgrignoli P, \textit{J. Stat. Mech.} P07021 (2010)

\bibitem{DGM08} Dorogovtsev SN, Goltsev AV, Mendes JFF, \textit{Rev. Mod. Phys.} \textbf{80}, 1275 (2008)

\bibitem{MPWZ02} Mulet R, Pagnani A, Weigt M, Zecchina R, \textit{Phys. Rev. Lett.} \textbf{89}, 268701 (2002)

\bibitem{ZK07} Zdeborov\'a L, Krzakala F, \textit{Phys. Rev. E} \textbf{76}, 031131 (2007)

\bibitem{FA79} Fishman S, Aharony A, \textit{J. Phys. C} \textbf{12}, L729 (1979)

\bibitem{C84} Cardy J, \textit{Phys. Rev. B} \textbf{29}, 505 (1984)

\bibitem{KRTZ10} Krzakala F, Ricci-Tersenghi F, Zdeborov\'a L, \textit{Phys. Rev. Lett.} \textbf{104}, 207208 (2010)

\bibitem{KRTSZ11} Krzakala F, Ricci-Tersenghi F, Sherrington D, Zdeborov\'a L, \textit{J. Phys. A} \textbf{44}, 042003 (2011)

\bibitem{dDM87} DeDominicis C, Mottishaw P, \textit{in} Lecture Notes in Physics, vol. 268, p.121 (1987)

\bibitem{dAT78} de Almeida J R L, Thouless D J, \textit{J. Phys. A} \textbf{11}, 983 (1978)

\bibitem{dA03} de Almeida J R L, \textit{Braz. J. Phys.} \textbf{33}, 892 (2003)

\end{thebibliography}
\end{document}